\documentclass[lnicst]{svmultln}
\usepackage{makeidx}  
\usepackage{graphicx}
\begin{document}
\mainmatter              
\title{Modeling multistage decision processes with Reflexive Game Theory}
\titlerunning{Modeling Mixed Groups of Humans and Robots} 
%
\author{Sergey Tarasenko}
\authorrunning{Sergey Tarasenko}   
\tocauthor{Sergey Tarasenko}

\institute{
\email{infra.core@gmail.com}}

\maketitle              

\begin{abstract}        
This paper introduces application of Reflexive Game Theory
to the matter of multistage decision making processes. The idea behind
is that each decision making session has certain parameters like ``when
the session is taking place", ``who are the group members to make decision",
``how group members influence on each other", etc. This study illustrates the
consecutive or sequential decision making process, which consist of two stages. During the stage 1 decisions about the parameters of the ultimate decision making are made. Then stage 2 is implementation of Ultimate decision making itself. Since during stage 1 there can be multiple decision sessions. In such a case it takes more than two sessions to make ultimate (final) decision. Therefore the overall process of ultimate decision making becomes multistage decision making process consisting of consecutive decision making sessions.
\keywords{Reflexive Game Theory (RGT), multi-stage decision making}
\end{abstract}

\section{Introduction}
The Reflexive Game Theory (RGT) \cite{lef2,lef5} allows to predict choices of subjects in the group. To do so, the information about a group structure and mutual influences between subjects is needed. Formulation and development of RGT was possible due to fundamental psychological research in the field of reflexion, which had been conducted by Vladimir Lefebvre \cite{lef4}.

The group structure means the set of pair-wise relationships between subjects in the group. These relationships can be either of alliance or conflict type. The mutual influences are formulated in terms of elements of Boolean algebra, which is built upon the set of universal actions. The elements of Boolean algebra represent all possible choices. The mutual influences are presented in the form of Influence matrix. 

In general, RGT inference can be presented as a sequence of the following steps \cite{lef2,lef5}:

1)	formalize choices in terms of elements of Boolean algebra of alternatives;

2)	presentation of a group in the form of a fully connected relationship graph, where solid-line and dashed-line ribs (edges) represent alliance and conflict relationships, respectively; 

3)	if relationship graph is decomposable, then it is represented in the form of polynomial: alliance and conflict are denoted by conjunction ($\cdot$) and disjunction (+) operations;

4)	diagonal form transformation (build diagonal form on the basis of the polynomial and fold this diagonal form);

5)	deduct the decision equations;

6)	input influence values into the decision equations for each subject.

Let us call the process of decision making in a group to be a session. Therefore, in RGT models a single session. 

\section{Model of two-stage decision making: formation of points of view}

This study is dedicated to the matter of setting mutual influences in a group by means of \textit{reflexive control} \cite{lef1}. 
The influences, which subjects make on each other, could be considered as a result of a decision making session previous to \textit{ultimate decision making (final session)}. We will call the influences, obtained as a result of a previous session(s), a \textit{set-up influences}. The set-up influences are intermediate result of the overall decision making process. The term set-up influences is related to the influences, which are used during the final session, only. 

\begin{figure}
\centering
\includegraphics[height=2cm]{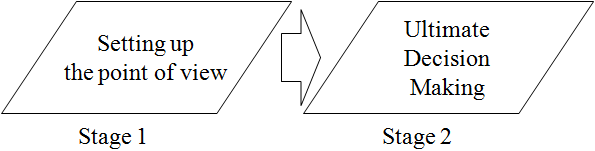}
\caption{The general schema of the two-stage decision making.}
\label{twostage}
\end{figure}

Consequently, the overall decision making process could be segregated into two stages.  Let the result of such discussion (decision making) be a particular decision regarding the matter under consideration. We assume the actual decision making regarding the matter (final session - Stage 2) is preceded by preliminary session (Stage 1). Stage 1 is about a decision making regarding the influences (points of view), which each subject will support during the final session. We call such overall decision making process to be a \textit{two-stage decision making process}. The general schema of a two-stage decision making is presented in Fig.\ref{twostage}.

To illustrate such model we consider a simple example.

\textit{Example 1.} Let director of some company has a meeting with his advisors. The goal of this meeting is to make decision about marketing policy for the next half a year. The background analysis and predictions of experts suggest three distinct strategies: aggressive (action $\alpha$), moderate (action $\beta$) and soft (action $\gamma$) strategies. 
The points of view of director and his advisors are formulated in terms of Boolean algebra of alternatives. Term point of view implies that a subject makes the same influences on the others. Director supports moderate strategy ($\{\alpha\}$), the 1st and the 2nd advisors are supporting aggressive strategy ($\{\beta\}$), and the 3rd advisor defends the idea of soft strategy ($\{\gamma\}$). The matrix of initial influences is presented in Table \ref{infMtx}.

\begin{table}
\caption{Matrix of initial points of view (influences) used in Example 1}
\begin{center}
\begin{tabular}{|c|c|c|c|c|}
\hline
{}&a&b&c&d\\
\hline
\rule{0pt}{12pt}a&a&$\{\alpha\}$&$\{\alpha\}$&$\{\alpha\}$\\[2pt]
\hline
\rule{0pt}{12pt}b&$\{\alpha\}$&b&$\{\alpha\}$&$\{\alpha\}$\\[2pt]
\hline
\rule{0pt}{12pt}c&$\{\beta\}$&$\{\beta\}$&c&$\{\beta\}$\\[2pt]
\hline
\rule{0pt}{12pt}c&$\{\gamma\}$&$\{\gamma\}$&$\{\gamma\}$&d\\[2pt]
\hline
\end{tabular}
\end{center}
\label{infMtx}
\end{table}

Let director is in a conflict with all his advisors, but his advisors are in alliance with each other. Variable $c$ represents the Director, variables $a$, $b$ and $d$ correspond to the 1st , the 2nd and the 3rd  advisor, respectively.

\begin{figure}
\centering
\includegraphics[height=2cm]{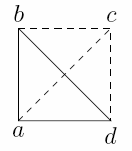}
\caption{Relationship graph for a director-advisors group.}
\label{polyn1}
\end{figure}

The relationship graph is presented in Fig.\ref{polyn1}. Polynomial $abd+c$ corresponds to this graph. 

After diagonal form transformation the polynomial does not change:

\[\begin{array}{*{20}{c}}
   {} & {} & {[a][b][d]} & {} & {} & {} & {} & {}  \\
   {} & {[abd] }  & {} &{+[c]} & {} & {} & {1 + [c]} & {}  \\
   {[abc+d]} & {} & {} & {} &  =  & {[abd+c]} & {} & { = abd+c.}  \\
\end{array}\]

Then we obtain four decision equations and their solutions (decision intervals) (Table \ref{decInt}).

\begin{table}
\caption{Decision intervals for Example 1}
\begin{center}
\begin{tabular}{|c|c|c|}
\hline
{Subject}&{Decision Equations}&{Decision Intervals}\\
\hline
\rule{0pt}{12pt}a&$a=(bd+c)a+c\overline{a}$&$(bd+c)\supseteq a \supseteq c$\\[2pt]
\hline
\rule{0pt}{12pt}b&$b=(ad+c)b+c\overline{b}$&$(ad+c)\supseteq b \supseteq c$\\[2pt]
\hline
\rule{0pt}{12pt}c&$c=c+abd\overline{c}$&$1\supseteq c \supseteq abd$\\[2pt]
\hline
\rule{0pt}{12pt}d&$d=(ab+c)d+c\overline{d}$&$(ab+c)\supseteq d \supseteq c$\\[2pt]
\hline
\end{tabular}
\end{center}
\label{decInt}
\end{table}

Next we calculate the decision intervals by using information from the influence matrix:

subject a:  $(bd+c)\supseteq a \supseteq c$ $\Rightarrow$ $(\{\alpha\} \{\gamma\}+\{\beta\})\supseteq a \supseteq \{\beta\}$ $\Rightarrow$ $a=\{\beta\}$;

subject b:  $(ad+c)\supseteq b \supseteq c$ $\Rightarrow$ $(\{\alpha\} \{\gamma\}+\{\beta\})\supseteq b \supseteq \{\beta\}$ $\Rightarrow$ $b=\{\beta\}$;

subject c:  $1\supseteq c \supseteq abd$$\Rightarrow$ $1\supseteq c \supseteq \{\alpha\}\{\alpha\}\{\gamma\} $ $\Rightarrow$ $1\supseteq c \supseteq 0 $ $\Rightarrow$ $c = c$;

subject d:  $(ab+c)\supseteq d \supseteq c$ $\Rightarrow$ $(\{\alpha\} c+\{\beta\})\supseteq b \supseteq \{\beta\}$ $\Rightarrow$ $\{\alpha, \beta\} \supseteq d \supseteq \{\beta\}$.

Therefore, after the preliminary sessions, the point of view of the subjects have changed. Director has obtained a freedom of choice, since he can choose any alternative: $1 \supseteq c \supseteq 0$ $\Rightarrow$ $c = c$. At the same time, the 1st and the 2nd advisors support moderate strategy ($a$ = $b$ = $\{\beta\}$). Finally, the 3rd advisor now can choose between points of view $\{\alpha,\beta\}$ (aggressive of moderate strategy) and $\{\beta\}$ (moderate strategy): $\{\alpha, \beta\} \supseteq d \supseteq \{\beta\}$.

Thus, the point of view of the 1st and the 2nd advisors is strictly determined, while the point of view of 3rd advisor is probabilistic. 

Next we calculate choice of each subject during the final session, considering the influences resulting from the preliminary session. The matrix of set-up influences is presented in Table \ref{setupInf}. The intervals in the matrix imply that a subject can choose either of alternatives from the given interval as a point of view. 

\begin{table}
\caption{The matrix of set-up influences for Example 1.}
\begin{center}
\begin{tabular}{|c|c|c|c|c|}
\hline
{}&a&b&c&d\\
\hline
\rule{0pt}{12pt}a&a&$\{\beta\}$&$\{\beta\}$&$\{\beta\}$\\[2pt]
\hline
\rule{0pt}{12pt}b&$\{\beta\}$&b&$\{\beta\}$&$\{\beta\}$\\[2pt]
\hline
\rule{0pt}{12pt}c&$1 \supseteq c \supseteq 0$ &$1 \supseteq c \supseteq 0$ &c&$1 \supseteq c \supseteq 0$\\[2pt]
\hline
\rule{0pt}{12pt}c& $\{\alpha, \beta\} \supseteq d \supseteq \{\beta\}$ & $\{\alpha, \beta\} \supseteq d \supseteq \{\beta\}$ & $\{\alpha, \beta\} \supseteq d \supseteq \{\beta\}$ &d\\[2pt]
\hline
\end{tabular}
\end{center}
\label{setupInf}
\end{table}

Subject $a$: 

$d=\{\alpha,\beta\}$: $(bd+c)\supseteq a \supseteq c$ $\Rightarrow$$(\{\beta\}\{\alpha,\beta\}+c)\supseteq a \supseteq c$ $\Rightarrow$$(\{\beta\}+c)\supseteq a \supseteq c$;

$d=\{\beta\}$: $(bd+c)\supseteq a \supseteq c$ $\Rightarrow$$(\{\beta\}\{\beta\}+c)\supseteq a \supseteq c$ $\Rightarrow$$(\{\beta\}+c)\supseteq a \supseteq c$;

Subject $b$: 

$d=\{\alpha,\beta\}$: $(ad+c)\supseteq b \supseteq c$ $\Rightarrow$$(\{\beta\}\{\alpha,\beta\}+c)\supseteq b \supseteq c$ $\Rightarrow$$(\{\beta\}+c)\supseteq b \supseteq c$;

$d=\{\beta\}$: $(ad+c)\supseteq b \supseteq c$ $\Rightarrow$$(\{\beta\}\{\beta\}+c)\supseteq b \supseteq c$ $\Rightarrow$$(\{\beta\}+c)\supseteq b \supseteq c$;

Subject $c$: 

$d=\{\alpha,\beta\}$: $1 \supseteq c \supseteq abd$ $\Rightarrow$$1 \supseteq c \supseteq \{\beta\}\{\beta\}\{\alpha,\beta\}$ $\Rightarrow$$1 \supseteq c \supseteq \{\beta\}$;

$d=\{\beta\}$: $1 \supseteq c \supseteq abd$ $\Rightarrow$$1 \supseteq c \supseteq \{\beta\}\{\beta\}\{\beta\}$ $\Rightarrow$$1 \supseteq c \supseteq \{\beta\}$;

Subject $d$: 

$(ab+c)\supseteq d \supseteq c$ $\Rightarrow$$(\{\beta\}\{\beta\}+c)\supseteq d \supseteq c$ $\Rightarrow$$(\{\beta\}+c)\supseteq d \supseteq c$.

Now we compare the results of a single session with the ones of the two-stage decision making. 

The single session case has been considered above. If the final decision has to be made after the single session, then the 3rd advisor would be able to choose alternative $\{\alpha,\beta\}$ and realize action $\alpha$. This option implies that each advisor is responsible for a particular part of the entire company and can take management decisions on his own.

Next we consider the decision made after the two-stage decision making. In such a case, regardless of influence of the 3rd advisor (subject $d$), choice of advisors $a$ and $b$ is defined by the interval $\{\beta\}+c \supseteq x \supseteq c$, where $x$ is either $a$ or $b$ variable. Thus, if director is inactive ($c=0$), subjects $a$ and $b$ can choose either moderate strategy ($\{\beta\}$) or make no decision ($0=\{\}$). The same is true for subject $d$.

If the director makes influence $\{\beta\}$, then all advisors will choose alternative $\{\beta\}$.

The director himself can choose from the interval $1 \supseteq c \supseteq \{\beta\}$ after the final session. This means that the director can choose any alternative, containing action $\beta$. Thus, occasionally the director can realize his initial point of view as moderate  strategy.

This example illustrates how using the two-stage decision making it is possible to make one’s opponents choose the one’s point of view. Meanwhile a person interested in such reflexive control can still sustain the initial point of view.
The obtained results are applicable in both cases when 1) only the director makes a decision; or 2) the decision are made individually by each subject.

\section{A Model of a multi-stage decision making: set-up parameters of the final session}

Now we consider the two-stage model in more details. In the considered example, during the preliminary session only the decision regarding the influences has been under consideration.  In general case, however, before the final session has begun, there can be made decisions regarding any parameters of the final session. Such parameters include but are not limit to:

1)	group structure (number of subjects and relationships between subjects in a group);

2)	points of view;

3)	decision to start a final session (a time when the final session should start), etc.

We call the decision regarding a single parameter to be \textit{consecutive decision}, and decisions regarding distinct parameters to be \textit{parallel decisions}. 

\begin{figure}
\centering
\includegraphics[height=3cm]{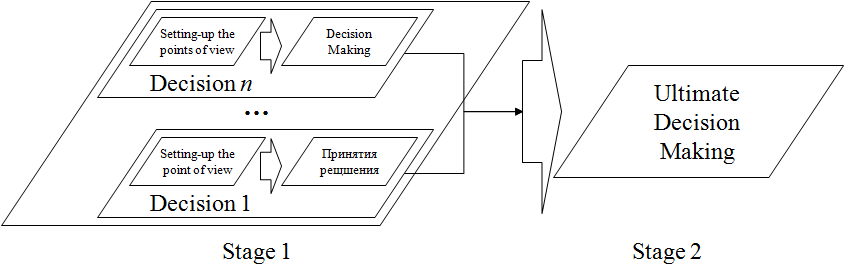}
\caption{Multi-stage decision making model.}
\label{multistage}
\end{figure}

Therefore, during the first stage (before the final session) it is possible to make multiple decisions regarding various parameters of the final session. This decisions could be both parallel and consecutive ones. We call such model of decision making to be a \textit{multi-stage process of decision making} (Fig.\ref{multistage}).

\section{Modeling multi-stage decision making processes with RGT}
Next we consider realization of multi-stage decision making with RGT. 

\textit{Example 2: Change a group structure.} Considering the subject from Example 1, we analyze the case when director wants to exclude the 3rd advisor from the group, which will make the final decision. 

\begin{figure}
\centering
\includegraphics[height=2cm]{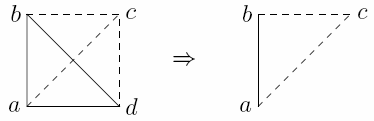}
\caption{Exclusion of a subject $d$ from a group.}
\label{polyn2}
\end{figure}

In such a case, there is a single action – 1 – to exclude subject $d$ from the group. Then Boolean algebra of alternatives includes only two elements: 1 and 0. Furthermore, it is enough that director just raise a question to exclude subject $d$ from a group and make influence 1 on each subject: if  $с = 1$, then $a=1$, $b=1$ and $d=1$ (Table \ref{decInt}). Thus the decision to exclude subject $d$ from the group would be made automatically (Fig.\ref{polyn2}).

\textit{Example 3: Realization of a multi-stage decision making.} Let the first decision discussed during the first stage is a decision regarding influences (points of view). The next decision was about exclusion of a subject $d$ from the group. Thus, during the first step the formation (setting-up) of points of view has been implemented, then the structure of a group was changed. Therefore the group, which should make a final decision is described by polynomial  $ab +c$. The decision equations and their solutions are presented in Table \ref{decInt3}.
 
\begin{figure}
\centering
\includegraphics[width=12cm]{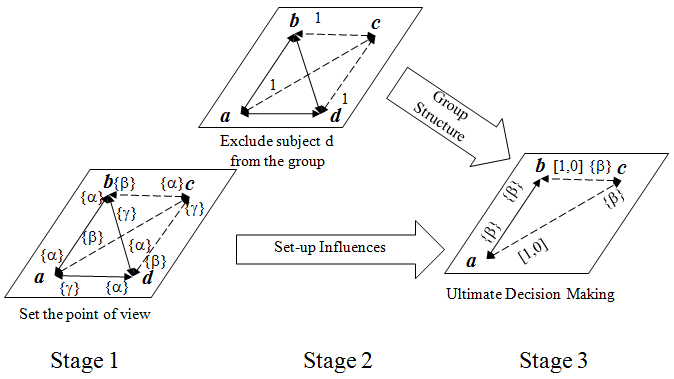}
\caption{Illustration of multi-stage decision making process. The influences are indicated by the arrow-ends of the ribs. The actual influence is presented near the arrow-end. }
\label{multiEx}
\end{figure}

The overall multi-stage decision making process is presented in Fig.\ref{multiEx}. 

\begin{table}
\caption{Decision intervals for Example 3}
\begin{center}
\begin{tabular}{|c|c|c|}
\hline
{Subject}&{Decision Equations}&{Decision Intervals}\\
\hline
\rule{0pt}{12pt}a&$a=(b+c)a+c\overline{a}$&$(b+c)\supseteq a \supseteq c$\\[2pt]
\hline
\rule{0pt}{12pt}b&$b=(a+c)b+c\overline{b}$&$(a+c)\supseteq b \supseteq c$\\[2pt]
\hline
\rule{0pt}{12pt}c&$c=c+ab\overline{c}$&$1\supseteq c \supseteq ab$\\[2pt]
\hline
\end{tabular}
\end{center}
\label{decInt3}
\end{table}

We consider that the point of view cannot change without preliminary session regarding the parameter. Therefore we assume that the points of view do not change after the change of group structure.

Therefore, during the final session the subjects would make the set-up influences derived from the preliminary session: subjects $a$ and $b$ will make influences $\{\beta\}$ and subject $c$ will have a choice from the interval $1 \supseteq c \supseteq \{\beta\}$. 

Such process is introduced in Fig. \ref{multiEx}. During the 1st stage (first step), the points of view of subjects have been formed. On the 2nd stage (second step), the decision to exclude subject $d$ from a group has been made. Finally, during the 3rd stage the final decision regarding the marketing strategy has been made.

\section{Discussion and conclusion}

This study introduces the two-stage and multi-stage decision making processed. During the last stage the final decision is made. During the earlier stages the decisions regarding the parameters of a final session are considered. 

This study shows how before the final decision making the intermediate decision regarding parameters of the final session can be made and how to overall process of decision making could be represented as a sequence of decision making sessions.

This approach enables complex decision making, which involves numerous parameters.

The important feature of the multi-stage decision making is that during the preliminary sessions subjects can convince other subjects to accept their own point of view. Therefore other subjects can be convince to make decisions beneficial for a particular one. Such approach also allows to distribute the responsibility between all the members of the group, who make the final decision.

The results presented in this study allow to extend the scope of applications of  RGT to modeling of multi-stage decision making processes. Therefore it becomes possible to perform scenario analysis of various variants of future trends and apply reflexive control to the management of projects.


\begin{thebibliography}{5}

\bibitem{lef2} Lefebvre, V.A.: Lectures on Reflexive Game Theory.
Cogito-Centre, Moscow (2009) (in Russian)

\bibitem{lef5} Lefebvre, V.A.: Lectures on Reflexive Game Theory.
Leaf \& Oaks, Los Angeles (2010)



\bibitem{lef4} Lefebvre, V.A.: Algebra of Conscience. 
D. Reidel, Holland (1982)

\bibitem{lef1} Lefebvre, V.A.: The basic ideas of reflexive game's logic. Problems of systems and structures research. 73--79 (1965) (in Russian)


\end{thebibliography}
\end{document}